\documentclass[pra,showpacs,nofootinbib]{revtex4}
\usepackage{hyperref}
\usepackage{amsmath}
\usepackage{amssymb}
\usepackage{amsthm}
\usepackage{graphics}
\usepackage{graphicx}
\usepackage{color}
\usepackage[usenames,dvipsnames,svgnames,table]{xcolor}
\usepackage{refcount}

\long\def\ca#1\cb{}

\def\bra#1{\langle#1|}
\def\dyad#1#2{|#1\rangle\langle#2|}

\def\ket#1{|#1\rangle }

\def\AC{{\cal A}}
\def\BC{{\cal B}}
\def\CC{{\cal C}}

\def\EC{{\cal E}}

\def\HC{{\cal H}}

\def\KC{{\cal K}}

\def\MC{{\cal M}}

\def\SC{{\cal S}}

\def\endproof{{\hspace{\stretch{1}}$\blacksquare$}}

\newtheorem{thm1}{Theorem}
\newtheorem{cor2}[thm1]{Corollary}
\newtheorem{cor5}[thm1]{Corollary}

\newtheorem{thm4}[thm1]{Theorem}

\newtheorem{thm102}[thm1]{Theorem}
\newtheorem{thm103}[thm1]{Theorem}
\newtheorem{thm104}[thm1]{Theorem}

\newtheorem{thm106}[thm1]{Theorem}

\newtheorem{lem1}[thm1]{Lemma}

\begin{document}
\title{Conditions for uniqueness of product representations for separable quantum channels and separable quantum states}
\author{Scott M. Cohen}
\email{cohensm52@gmail.com}
\affiliation{Department of Physics, Portland State University,
Portland, Oregon 97201}

\begin{abstract}
We give a sufficient condition that an operator sum representation of a separable quantum channel in terms of product operators is the unique product representation for that channel, and then provide examples of such channels for any number of parties. This result has implications for efforts to determine whether or not a given separable channel can be exactly implemented by local operations and classical communication. By the Choi-Jamiolkowski isomorphism, it also translates to a condition for the uniqueness of product state ensembles representing a given quantum state. These ideas follow from considerations concerning whether or not a subspace spanned by a given set of product operators contains at least one additional product operator.
\end{abstract}

\date{\today}

\maketitle
\section{Introduction}\label{sec1}
The evolution of quantum systems is a subject of intense interest, studied in a wide range of contexts. One example of such studies is a question of prime importance in quantum information theory, to find the amount of information that can be carried by a system that is transmitted through a noisy quantum channel. The noise in the channel may be thought of as arising from interaction of the system with an environment. In general terms, such an evolution can be described \cite{Kraus} by the completely positive, trace-preserving map,
\begin{align}\label{eqn1}
\EC(\rho) = \sum_jK_j\rho K_j^\dag,
\end{align}
where $\rho$ is the initial state of the system, $\EC(\rho)$ the state following the evolution, and the set of Kraus operators $\{K_j\}$ is constrained only by the condition,
\begin{align}\label{eqn2}
I=\sum_jK_j^\dag K_j,
\end{align}
which guarantees that $\EC$ is trace-preserving. Eq.~\eqref{eqn1} is known as an operator-sum representation of $\EC$.

It is well-known that the description of a quantum channel in terms of Kraus operators, $\{K_j\}$, is not unique \cite{NielsenChuang}. Any other set $\{K_i^\prime\}$ that satisfies
\begin{align}\label{eqn3}
K_i^\prime = \sum_ju_{ij}K_j,
\end{align}
with matrix $u$ an isometry satisfying $\sum_i u_{ij}^\ast u_{ik}=\delta_{jk}$, is a valid representation of the exact same channel as that represented by the set $\{K_j\}$.\footnote{\label{ftnt1}If the original set is linearly dependent, it is possible for the new set to have fewer members, in which case $u$ would not be an isometry. By padding the new set with zero operators, $u$ can always be extended to form a unitary matrix.}

Another important area of study in quantum information theory is the so-called `distant-labs' paradigm. In this scenario, two or more spatially separated parties each hold quantum subsystems, and one wishes to understand how the global system evolves under the actions of these parties. Often, one considers that each of the $P$ parties can act locally on their own subsystem, and they can all exchange classical information with one another, a process collectively known as local operations and classical communication, or LOCC. Understanding what can be accomplished under these conditions is an extremely challenging problem, and it has often been helpful to study a strictly larger class of actions, that known as separable operations \cite{Rains}. A separable operation is a type of quantum channel, and each `separable channel' is characterized by the fact that there exists at least one set of Kraus operators representing that channel in which every Kraus operator is a product operator, of the form $K^{(1)}\otimes K^{(2)}\otimes\ldots\otimes K^{(P)}$. An LOCC protocol necessarily implements a set of Kraus operators that are all product operators, so if a channel can be implemented by LOCC, it must be separable. However, there exist separable channels that cannot be implemented by LOCC \cite{Bennett9}.

In a series of recent papers \cite{mySEPvsLOCC,myMany,myExtViolate}, we have studied the conditions under which a given separable measurement can be implemented by LOCC. In \cite{mySEPvsLOCC, myMany}, we presented a method of constructing a finite-round LOCC protocol for a given separable measurement whenever such a protocol exists. Our meaning, here, when we talk about a separable \emph{measurement}, is to refer to a specific set of Kraus operators (up to multiplicative factors), so these results do not automatically answer the question of when a separable \emph{channel} can be implemented by LOCC. The reason is that, because of the fact there are many possible sets of Kraus operators that represent the same quantum channel, showing that any one set is not LOCC leaves open the possibility that another set may be. Therefore, in order for one to use the method of \cite{mySEPvsLOCC, myMany} to say with certainty that a given channel cannot be implemented by LOCC, one has to check every possible set of Kraus operators that represents that channel. 

Of course, since an LOCC protocol implements only product Kraus operators, one really only has to check all possible product representations of a given separable channel. Nonetheless, even given a single product representation for a given channel, it is not at all clear how to find one other such representation, let alone \emph{all} others. It has not even been clear how to answer the question of the \textit{existence} of other product representations, and this is true no matter how many product representations are already known for a given channel. Therefore, it appears to be a very difficult challenge to find conditions that could be applied in general, capable of determining that a separable channel (as opposed to a separable measurement) cannot be implemented by LOCC.

In the next section, we use a result of \cite{ShengjunSumProd} to make a first step toward answering this challenge by giving a general condition that a product representation is the unique product representation for a given separable channel. Given a unique product representation, one can use the methods of \cite{mySEPvsLOCC,myMany} to check that single representation to see if it can be exactly implemented by LOCC, thereby directly determining whether or not the channel itself can be so implemented. In addition, we provide examples of separable channels that have unique product representations, some of which can be implemented by LOCC, while others cannot. The importance of these examples is that, at least as far as we are aware, it has not previously been known whether or not there exist quantum channels for which there is one, and only one, product Kraus representation. Therefore, these examples provide some insight into the structure of separable channels and LOCC. In Section~\ref{sec3}, we invoke the Choi-Jamiolkowski isomorphism \cite{Choi,Jamiolkowski} to apply these ideas to the task of determining when a representation of a separable quantum state by an ensemble of product pure states is the unique such product representation, a question of independent interest that has arisen previously in the context of studying the facial structure of the convex set of all separable states \cite{AlfsenShultz,Hyeok}. In Section~\ref{sec4} we present a related result indicating when a set of product operators can represent a complete quantum measurement. In the appendix, we present results (which are used to derive those just mentioned) about when the space spanned by a set of product operators can contain an additional product operator.

\section{Uniqueness of product representations for separable quantum channels}\label{sec2}
The results of this and subsequent sections are based on the following theorem that relates to whether, by considering linear combinations of a fixed set of product operators, one can find a product operator that is not in that original set.
\begin{thm106}\label{thm106}Given a set of product operators for $P$ parties, $\{\MC_j=M_j^{(1)}\otimes\ldots\otimes M_j^{(P)}\}_{j=1}^N$, if there exists a set of nonzero coefficients, $\{c_j\}$, such that the linear combination $\SC=\sum_{j=1}^Nc_j\MC_j$ is a product operator, then for each pair of parties $\alpha,\beta$,
\begin{align}\label{eqn992}
\delta_\alpha+\delta_\beta\le N+1,
\end{align}
where $\delta_\alpha$ is the dimension of the space spanned by operators $\{M_j^{(\alpha)}\}_{j=1}^N$.
\end{thm106}
\noindent This result was found previously, appearing as proposition $3$ in \cite{ShengjunSumProd} (that proposition is stated there in terms of linear combinations of product pure states instead of product operators, but see Section \ref{sec3} for a discussion of the close relationship between these contexts). For completeness, we provide a proof in the appendix, where various related results concerning linear combinations of product operators are also proved.

Using this theorem, we can give a condition under which a given representation of a separable quantum channel, in terms of product Kraus operators, is the unique product representation. If the channel is separable, there exists a set $\{K_j\}$ that are all product operators. Then, for there to exist a different product representation for that channel, the $K_i^\prime$ in \eqref{eqn3} must also be product for each $i$. Obviously, if the set $\{K_i^\prime\}$ is to differ from the set $\{K_j\}$, then either (a) the former is a proper subset of the latter (up to multiplicative factors), or (b) (at least) one of the $K_i^\prime$ must be different from every one of the $K_j$. In the latter case, that $K_i^\prime$ must lie in the span of $\{K_j\}$, by \eqref{eqn3}. In the first case, each $K_j$ in the complement of $\{K_i^\prime\}$ within $\{K_j\}$ must be a linear combination of the $K_i^\prime$ (just swap their respective roles in \eqref{eqn3}), implying directly that every operator in that complement lies in the span of $\{K_i^\prime\}\subset\{K_j\}$. Therefore, if there is a product representation of the channel that differs from $\{K_j\}$, there must be a subset containing $n\ge 2$ of the $K_j$ whose span contains a product operator that is not a member of that subset. Given this, there is then a smallest subset whose span contains that product operator, and this smallest subset must satisfy the conditions of Theorem~\ref{thm106}, with coefficients $c_j$ all non-zero. That is, for some such subset, we must have that for every pair of parties $\alpha,\beta$,
\begin{align}\label{eqn402}
\delta_\alpha+\delta_\beta\le n+1.
\end{align}
Thus, we have
\begin{thm103}\label{thm103}
If no subset of the product Kraus operators $\{K_j\}$ containing $n\ge2$ members exists such that \eqref{eqn402} is satisfied for every pair of parties $\alpha,\beta$, then that set is the unique set of product Kraus operators representing the given separable quantum channel.
\end{thm103}
\noindent Note that the $\delta_\alpha$ will be different for different subsets, so it will be necessary to check all subsets. Notice also that the conditions of this theorem imply that the full set of Kraus operators is linearly independent. However, linear independence is not by itself a strong enough condition, as there exist linearly independent sets of product Kraus operators that are not the unique product representations for the given channel. A simple example illustrating the latter point is a projective measurement on two qubits in the standard basis, consisting of Kraus operators $\{\ket{00}\bra{00},\ket{01}\bra{01},\ket{10}\bra{10},\ket{11}\bra{11}\}$. Any (isometric) set of coefficients $u_{ij}$ in \eqref{eqn3}, which mixes only the first pair of operators and/or only the second pair, yields another product representation for the same quantum channel. Of course, these do not exhaust the full set of product representations, but they are enough to illustrate the point, demonstrating that, in fact, there exist product representations that are linearly independent for which there is an uncountable infinity of other, distinct, product sets representing the same quantum channel.

Consider a separable channel on two qubits having the following representation in terms of product Kraus operators,
\begin{align}\label{eqn701}
K_1&=\dyad{0}{1}\otimes\left(\dyad{0}{0}+e^{\textrm{i}\phi}\sqrt{1-|\mu|^2}\dyad{1}{1}\right)\notag\\
K_2&=\left(\dyad{0}{0}+\mu\dyad{1}{1}\right)\otimes\dyad{0}{1}\notag\\
K_3&=\dyad{1}{0}\otimes\dyad{1}{0}.
\end{align}
Can this channel be implemented by LOCC? The method of \cite{mySEPvsLOCC} for constructing LOCC protocols from sets of product Kraus operators considers the local parts of positive operators, $\KC_j:=K_j^\dag K_j$,
\begin{align}\label{eqn702}
\KC_1&=\dyad{1}{1}\otimes\left(\dyad{0}{0}+(1-|\mu|^2)\dyad{1}{1}\right)\notag\\
\KC_2&=\left(\dyad{0}{0}+|\mu|^2\dyad{1}{1}\right)\otimes\dyad{1}{1}\notag\\
\KC_3&=\dyad{0}{0}\otimes\dyad{0}{0}.
\end{align}
As no two of the local operators for either party in \eqref{eqn702} are proportional to each other, it follows immediately from the method of \cite{mySEPvsLOCC} that the set of Kraus operators in \eqref{eqn701} cannot be exactly implemented by LOCC. We note that there had been no previously known method of determining if there are other representations of this channel in terms of product operators. However, since $\{\dyad{0}{1},\dyad{1}{0},\dyad{0}{0}+\nu\dyad{1}{1}\}$ is a linearly independent set for every $\nu$, we have that for every subset of $n\ge2$ of these Kraus operators, $\delta_\alpha=n~\forall{\alpha}$, implying $\delta_\alpha+\delta_\beta=2n>n+1$, violating \eqref{eqn402}. Hence, the set of product Kraus operators given in \eqref{eqn701} satisfies the conditions of Theorem~\ref{thm103}, and this is therefore the unique set of product Kraus operators representing the corresponding quantum channel. We may thus conclude that, indeed, this channel cannot be exactly implemented by LOCC.

Here is another example, which we have used elsewhere \cite{myExtViolate} to demonstrate a very strong difference between LOCC and separable channels. It is a class of examples, actually, providing cases for any number of parties. Define positive operators $\KC_j=\dyad{\Psi_j}{\Psi_j},~j=1,\ldots,N$, acting on $P$ parties, where $\ket{\Psi_j}=(D/N)^{1/2}\ket{\psi_j^{(1)}}\otimes\ldots\otimes\ket{\psi_j^{(P)}}$, $D=d_1d_2\ldots d_P$, $d_\alpha$ is the dimension of Hilbert space $\HC_\alpha$, with parties ordered such that $d_1\le d_2\le\ldots\le d_P$. The state on party $\alpha$'s subsystem is defined as
\begin{align}\label{eqn555}
\ket{\psi_j^{(\alpha)}}=\frac{1}{\sqrt{d_\alpha}}\sum_{m_\alpha=1}^{d_\alpha}e^{2\pi \textrm{i}jp_\alpha m_\alpha/N}\ket{m_\alpha}.
\end{align}
Here, $p_1=1$ and for $\alpha\ge2$, $p_\alpha=d_1d_2\ldots d_{\alpha-1}$, $\ket{m_\alpha}$ is the standard basis for party $\alpha$, and $N$ is chosen as any prime number exceeding $D$. Next, define Kraus operators
\begin{align}\label{eqn556}
K_j=\ket{\Phi_j}\bra{\Psi_j}, 
\end{align}
\noindent with normalized states $\ket{\Phi_j}=\ket{\phi_j^{(1)}}\otimes\ket{\Phi_j^\prime}$ and $\{\ket{\Phi_j^\prime}\}_{j=1}^N$ a set of linearly independent product states on the $P-1$ parties excluding party $1$ (this requires that at least one output dimension of those last $P-1$ parties exceeds its input, in order that the overall output dimension is not less than $N$, the number of these independent states). Then, since no two of the $\ket{\psi_j^{(1)}}$ are proportional to each other, and considering a bipartite split between party $1$ and all the rest (`party' $\overline 1$), we have that for every subset of the Kraus operators given in \eqref{eqn556} having $n\ge2$ members, $\delta_1\ge2$ and $\delta_{\overline1}=n$ implying $\delta_1+\delta_{\overline 1}\ge n+2$. Therefore, the conditions of Theorem~\ref{thm103} are met for this set of Kraus operators, which is thus the unique product representation for the given channel. By the results of \cite{myExtViolate}, it cannot be exactly implemented by LOCC.

Other examples of separable quantum channels that have a unique representation in terms of product Kraus operators are very easy to construct. One such example has every Kraus operator proportional to a product unitary on $P\ge2$ parties, $K_j=\sqrt{q_j}U_j^1\otimes U_j^2\otimes\ldots\otimes U_j^P$, where $\sum_jq_j=1$ and each set of local unitaries, $\{U_j^\alpha\}_{j=1}^N$, is a linearly independent set. As a consequence of this linear independence, we see that every subset of $n\ge2$ of these Kraus operators violates \eqref{eqn402}, and this is thus a unique product Kraus representation for the given channel. We note, on the other hand, that this violation is much stronger than necessary, suggesting it may not be too difficult to construct examples for which none of the local operator sets are linearly independent. As an example which is valid for every $P\ge2$, consider a channel represented by $N=4$ product Kraus operators, which may (or may not) be unitaries as in the previous example, and such that there are at least two parties $\alpha,\beta$ having local operator sets that satisfy the following two conditions: (1) the span of every two local operators within each of these two sets is two-dimensional, and (2) the span of every three or more local operators within each of these two sets is three-dimensional. Then, $\delta_\alpha+\delta_\beta> n+1$ for every subset of more than one local operator, and this representation is therefore unique. A specific example is a two-qubit channel with Kraus operators proportional to product unitaries, $I\otimes I$, $\sigma_x\otimes\sigma_x, \sigma_y\otimes\sigma_y$, and $(I+i\sigma_x+i\sigma_y)\otimes(I+i\sigma_x+i\sigma_y)$. This channel can be implemented by LOCC using one round of communication.

Additional examples can be easily found. Starting from any separable channel, one can construct others that have unique representations in terms of product Kraus operators. One way to do this is as follows: start with a separable quantum channel represented by a set of $N$ Kraus operators $K_j$, which are product operators acting on $P$ parties. Choose two sets of $N$ linearly independent unitaries acting on two other parties, $\{U_j^{(P+1)}\}$ and $\{U_j^{(P+2)}\}$, and create a new set of Kraus operators defined as $\widetilde K_j=K_j\otimes U_j^{(P+1)}\otimes U_j^{(P+2)}$. The set $\{\widetilde K_j\}_{j=1}^N$ is a product representation of a separable channel on $P+2$ parties. From Theorem~\ref{thm103}, we see that Kraus operators $\{\widetilde K_j\}$ are the unique product representation of this new channel. Since we started with an arbitrary quantum channel, this provides a very general method of constructing channels, on any number of parties, that have unique product Kraus representations. The new channel can be implemented by LOCC if and only if the original set of Kraus operators can be so implemented, with the last two parties simply performing their unitaries at the end of any protocol that implements the original one, if one exists (implementing the original channel by another set of Kraus operators is not sufficient, as the parties will not then be able to perform each of the $U_j^{(\alpha)}$ for appropriate outcomes of the original protocol).

\section{Uniqueness of product state ensembles for separable quantum states}\label{sec3}
By the Choi-Jamiolkowski isomorphism \cite{Jamiolkowski,Choi} between quantum states and quantum channels, the results of the previous section can be carried over directly to quantum states. Instead of representations of quantum channels by sets of Kraus operators, quantum states are represented by ensembles of pure states, usually denoted by kets. Here, again, the representation is not unique, and the isometric freedom of representations of quantum channels as given in \eqref{eqn3}, becomes an equivalent isometric freedom of ensembles representing a quantum state. That is, a quantum state (a density operator) $\rho$ is represented by an ensemble of non-normalized pure states, $\{\ket{\Psi_j}\}$,
\begin{align}\label{eqn501}
\rho=\sum_j\dyad{\Psi_j}{\Psi_j}.
\end{align}
The freedom of representation tells us that any other ensemble, $\{\ket{\Psi_i^\prime}\}$, that represents this same state is related to the original one by
\begin{align}\label{eqn502}
\ket{\Psi_i^\prime}=\sum_{j=1}^Nu_{ij}\ket{\Psi_j},
\end{align}
where once again, coefficients $u_{ij}$ constitute the elements of an isometry, $\sum_iu_{ij}u_{ik}^\ast=\delta_{jk}$, compare \eqref{eqn3}.\footnotemark[\getrefnumber{ftnt1}]

The quantum state is separable if and only if there is an ensemble of pure product states that represents it, $\ket{\Psi_j}=\otimes_\alpha\ket{\psi_j^{(\alpha)}}$. As a consequence, all the proofs that we have given for linear combinations of product operators work just as well for states, since the basis of those proofs, given in the appendix as the proof of Theorem~\ref{thm1}, begins by reshaping those operators (in matrix representation) into column vectors, and those column vectors can just as well correspond to the kets that represent the product pure states of an ensemble for $\rho$. Then, recognizing \eqref{eqn502} as a linear combination of product kets equal to a product ket (compare to \eqref{eqn3} as a linear combination of product operators equal to a product operator), we have
\begin{thm104}\label{thm104}
If in a product ensemble $\{\ket{\Psi_j}\}$ representing separable state $\rho=\sum_j\dyad{\Psi_j}{\Psi_j}$ on any number of parties, no subset of the product pure states $\ket{\Psi_j}$ containing $n\ge2$ members exists such that \eqref{eqn402} is satisfied for every pair of parties $\alpha,\beta$, then that ensemble is the unique product ensemble representing the given separable quantum state. Note that here, $\delta_\alpha$ is the dimension of the span of the local kets $\{\ket{\psi_j^{(\alpha)}}\}_j$ for party $\alpha$ of the product states $\ket{\Psi_j}$ in the given subset.
\end{thm104}
\noindent This result generalizes earlier ones \cite{AlfsenShultz,KKirkpatrick,Hyeok}, where a special case of this theorem was proven for bipartite states and representations in which one of the two sets of local parts $\{\ket{\psi_j^{(\alpha)}}\}_j$ is linearly independent. Here, our theorem is valid  for any number of parties, and as is shown above for the case of quantum channels, it encompasses ensembles for which no local set is linearly independent (e.g., the two-qubit example for channels, given at the end of the second-to-last paragraph of the previous section, can be carried over to this case of quantum states). Furthermore, from the discussion around \eqref{eqn555} and \eqref{eqn556}, and also that given in the last paragraph of the previous section, we have methods of constructing examples of separable states on any number of parties for which there is a unique product ensemble representation.

\section{When a set of product operators can represent a quantum channel}\label{sec4}
Consider a bipartite LOCC protocol that implements Kraus operators $K_j=A_j\otimes B_j$ corresponding to positive operators $\KC_j=K_j^\dag K_j=\AC_j\otimes \BC_j$, with $\AC_j=A_j^\dag A_j$ and $\BC_j=B_j^\dag B_j$. Since this must be a (complete) separable measurement, we require that
\begin{align}\label{eqn201}
I_A\otimes I_B=\sum_{i=1}^{N} \AC_j\otimes\BC_j,
\end{align}
with $I_A,~I_B$ the identity operators on the respective party's Hilbert space. From the results of \cite{mySEPvsLOCC}, it is straightforward to argue that if the $\AC_j$ are linearly independent, then in order that an LOCC protocol for this set of operators exists, it must be that every $\BC_j$ is proportional to $I_B$. Then, $\delta_A=N,~\delta_B=1$, and $\delta_A+\delta_B=N+1$. In fact this holds true for all complete separable measurements, not just LOCC. Theorem~\ref{thm106} generalizes this idea to any number of parties, providing the following necessary condition on the positive operators corresponding to a separable measurement (or channel) on any multipartite system.
\begin{thm102}\label{thm102}
If a set of product operators $\{K_j=\otimes_{\alpha=1}^PK_j^{(\alpha)}\}_{j=1}^N$ on $P$ parties constitutes a Kraus represention of a separable channel, then it must be that dimensions $\delta_\alpha$ of the spans of local operator sets $\{K_j^{(\alpha) \dag}K_j^{(\alpha)}\}_{j=1}^N$ satisfy $\delta_\alpha+\delta_\beta\le N+1$ for each pair of parties $\alpha,\beta$.
\end{thm102}

\section{Conclusions}\label{conc}
In summary, we have presented a sufficient condition, Theorem~\ref{thm103}, that a set of product Kraus operators, representing a separable quantum channel $\EC$, is the unique product representation for $\EC$. When this condition is satisfied by a given set, one can use the method of \cite{mySEPvsLOCC,myMany} to directly determine whether or not this channel can be exactly implemented by LOCC. In addition we provided examples, which to our knowledge represent the first known cases of separable quantum channels that have a unique product Kraus representation. We then used the Choi-Jamiolkowski isomorphism to obtain a sufficient condition, Theorem~\ref{thm104}, for the uniqueness of a product state ensemble representing a given quantum state. We then gave a necessary condition for when a set of product Kraus operators can represent a complete quantum measurement.

The ideas underlying these results involve conditions under which there exists at least one linear combination of a given set of product operators that is a product operator. Our results for more than two parties are restricted to merely applying the bipartite result to every pair of parties, and it is natural to wonder if there exists a stricter characterization of sets of product operators for which there exist linear combinations that are product operators in the case of many parties. One can show that without imposing further constraints on the set of product operators, then it is definitely not possible to find a stricter characterization. This is seen from the following (admittedly rather contrived) example,
\begin{align}\label{eqn970}
\SC=\SC+\MC_1-\MC_1+\MC_2-\MC_2+\ldots+\MC_n-\MC_n,
\end{align}
where $n\in\mathbb{N}$, $\SC=S^{(1)}\otimes\ldots\otimes S^{(P)}$, $\MC_j=M_j^{(1)}\otimes\ldots\otimes M_j^{(P)}$, and local operator sets $\{S^{(\alpha)},\{M_j^{(\alpha)}\}_{j=1}^n\}$ are linearly independent for each $\alpha$. Then, $N=2n+1$, $\delta_\alpha=n+1~\forall{\alpha}$, and for every pair of parties $\alpha,\beta$ our upper bound is tight: $\delta_\alpha+\delta_\beta=2(n+1)=N+1$. Note that this leads to $\sum_\alpha\delta_\alpha=P(N+1)/2$, saturating another bound that can be directly obtained from the bipartite results. We do not know if a stricter characterization is possible for $P>2$ even by only excluding cases such as this that have proper subsets of the $N$ product operators for which there is a linear combination that vanishes (or equivalently, by requiring the $N$ product operators be linearly independent). We suspect, however, that if this is done the bound of $\delta_\alpha+\delta_\beta\le N+1$ may be far from tight for at least one, and likely for many, of the $\alpha,\beta$ pairs. It does appear that in many cases, including all our preliminary efforts with random numerical searches that require linear independence of the $N$ operators, the far stricter bound of $\sum_\alpha\delta_\alpha\le N+P-1$ is obeyed, and it would be interesting to know if there are general conditions under which this much stricter bound is valid.

\noindent\textit{Acknowledgments} --- I thank Vlad Gheorghiu and Li Yu for helpful comments. This work was supported in part by the National Science Foundation through Grant No. PHY-1205931.

\appendix*
\section{On linear combinations of product operators}\label{app1}
The main focus of this appendix is to study conditions under which there exists a linear combination of a fixed set of product operators that is itself a product operator. Note, however, that we are not just asking whether a given operator, obtained as a fixed sum of a set of product operators, is itself a product operator, but rather we are allowing the coefficients multiplying each operator in that sum to vary. As such, we are instead addressing the question of whether or not there exists an \emph{additional} product operator lying in the space spanned by a given set of product operators. Therefore the Schmidt rank --- which is only concerned with a single linear combination of product operators, where the coefficients in that linear combination are fixed --- does not resolve the questions we address here, which in many respects are considerably more involved. 

To gain some intuition into this question, consider an arbitrary linear combination of product operators,
\begin{align}\label{eqn60}
\SC=\sum_{j=1}^{N}c_jA_j\otimes B_j,
\end{align}
and assume that each term in the sum is non-vanishing. Regardless of the value of the coefficients, $c_j$, this is clearly a product operator if all the $A_j$ are proportional to each other, $A_j=a_jA_1$, such that
\begin{align}\label{eqn61}
\SC=A_1\otimes\sum_{j=1}^{N}c_ja_jB_j.
\end{align}
Notice that in this case the set $\{A_j\}_{j=1}^N$ spans a one-dimensional space, so that if $\delta_A$ is the dimension of the span of these operators, with a similar definition for $\delta_B$, then $\delta_A+\delta_B=\delta_B+1\le N+1$.

What happens if $\{A_j\}_{j=1}^N$ spans a two-dimensional space? If, for example, $A_1$ and $A_2$ are linearly independent, then $A_j=a_jA_1+a_j^\prime A_2$ and
\begin{align}\label{eqn62}
\SC=A_1\otimes\sum_{j=1}^{N}c_ja_jB_j+A_2\otimes\sum_{j=1}^{N}c_ja_j^\prime B_j.
\end{align}
This is a product operator if and only if $\sum_{j=1}^{N}c_ja_jB_j$ is proportional to $\sum_{j=1}^{N}c_ja_j^\prime B_j$. Given that sets $\{a_j\}$ and $\{a^\prime_j\}$ are strictly different ($a_1=1$, $a_2=0$, $a^\prime_1=0$, $a^\prime_2=1$), this is a non-trivial constraint on the $B$'s, implying the $\{B_j\}_{j=1}^N$ cannot span a space of dimension greater than $N-1$. Therefore, we again have that $\delta_A+\delta_B\le 2+N-1=N+1$. Under this condition, one can find non-zero coefficients $c_j$ such that $\SC$ is a product operator.

In the following, amongst other results, we will show that quite generally,
\begin{align}\label{eq63}
\delta_A+\delta_B\le N+1,
\end{align}
and we will also generalize this result to any number of parties.

Recall that the Schmidt rank  \cite{Tyson} of an operator $\SC$ is the smallest possible number of product operators that can be summed to obtain $\SC$. We begin with
\begin{thm1}\label{thm1}If there exists a set of nonzero coefficients, $\{c_j\}$, such that the linear combination
\begin{align}\label{eqn89}
\SC=\sum_{j=1}^Nc_jA_j\otimes B_j
\end{align}
has Schmidt rank $r_s$, then
\begin{align}\label{eqn90}
\delta_A+\delta_B\le N+r_s,
\end{align}
where $\delta_A$ is the dimension of the space spanned by operators $\{A_j\}_{j=1}^N$, and similarly for $\delta_B$.
\end{thm1}
\proof Reshape each operator $A_j$ (in matrix representation) into a column vector by stacking successive columns of $A_j$ each on top of the next one. Collect these column vectors together to form the columns of a new $d_A^2$-by-$N$ matrix, $\AC$, and do the same thing with the $B_j$ to form the $d_B^2$-by-$N$ matrix $\BC$. Define diagonal $N$-by-$N$ matrix $\CC$ with diagonal elements $c_j$, and denote the reshaped version of $S$ (each $d_B$-by-$d_B$ block of $S$ becomes one of the columns of $\SC$) as the $d_B^2$-by-$d_A^2$ matrix $\SC$. Then \eqref{eqn89} takes the reshaped form,
\begin{align}\label{eqn101}
\SC=\BC\CC\AC^T,
\end{align}
where $\AC^T$ is the transpose of $\AC$. Noting that the rank $\textrm{Rn}(XY)$ of a matrix product $XY$ satisfies the inequality \cite{HornJohnson},
\begin{align}\label{eqn102}
\textrm{Rn}(XY)\ge \textrm{Rn}(X)+\textrm{Rn}(Y)-l,
\end{align}
where $l$ is the inner dimension of the matrices ($X$ is $m$-by-$l$ and $Y$ is $l$-by-$n$), we see that
\begin{align}\label{eqn103}
\textrm{Rn}(\SC)\ge \textrm{Rn}(\BC)+\textrm{Rn}(\AC)-N,
\end{align}
where we have used the fact that $\textrm{Rn}(\CC)=N$, since the $c_j$ are assumed to be nonzero. Now, $\textrm{Rn}(\AC)$ is the number of linearly independent (reshaped) operators $A_j$, which is just $\delta_A$, and similarly, $\textrm{Rn}(\BC)=\delta_B$.

We next argue that $\textrm{Rn}(\SC)$ is equal to the Schmidt rank of $S$. To see this, write
\begin{align}\label{eqn104}
S=\sum_{k,l,m,n}\left(\sum_{j=1}^Nc_j(B_j)_{kl}(A_j)_{mn}\right)\dyad{km}{ln}=\sum_{k,l,m,n}S_{klmn}\dyad{km}{ln},
\end{align}
with $\ket{km}=\ket{k}_B\ket{m}_A$. The reshaped version of this is obtained by turning the bra $\bra{l}_B$ into a ket and the ket $\ket{m}_A$ into a bra. Then,
\begin{align}\label{eqn105}
\SC=\sum_{k,l,m,n}\left(\sum_{j=1}^Nc_j(B_j)_{kl}(A_j)_{mn}\right)\dyad{kl}{mn}=\sum_{k,l,m,n}S_{klmn}\dyad{kl}{mn}=\sum_{i=1}^{\textrm{Rn}(\SC)}\left(\sum_{kl}(b_i)_{kl}\ket{kl}\right)\left(\sum_{mn}(a_i)_{mn}\bra{mn}\right),
\end{align}
where the last equality follows from the definition of rank, with sets of vectors $\{a_i\}$ and $\{b_i\}$ each spanning a space of dimension $\textrm{Rn}(\SC)$. Identify $S_{klmn}=\sum_i(a_i)_{mn}(b_i)_{kl}$ and insert this into \eqref{eqn104} to obtain
\begin{align}\label{eqn106}
S=\sum_{i=1}^{\textrm{Rn}(\SC)}\sum_{k,l,m,n}\left[(a_i)_{mn}\dyad{m}{n}\right]\otimes\left[(b_i)_{kl}\dyad{k}{l}\right].
\end{align}
Noting the dimension of the spans of each set, $\{a_i\}$ and $\{b_i\}$, is $\textrm{Rn}(\SC)$, it is not difficult to show that $\textrm{Rn}(\SC)$ is the smallest number of product vectors by which $S$ can be expanded, from which we see that indeed, $\textrm{Rn}(\SC)$ is the Schmidt rank of $S$. Hence from \eqref{eqn103}, we have the desired result, $\delta_A+\delta_B\le N+r_s$.\hspace{\stretch{1}}$\blacksquare$

Setting $r_s=1$, we have the immediate corollary,
\begin{cor2}\label{cor2}If there exists a set of nonzero coefficients, $\{c_j\}$, such that the linear combination
\begin{align}\label{eqn91}
\SC=\sum_{j=1}^Nc_jA_j\otimes B_j
\end{align}
has Schmidt rank $1$ and so is a product operator, then
\begin{align}\label{eqn92}
\delta_A+\delta_B\le N+1.
\end{align}
\end{cor2}
We can generalize this result to any number of parties by simply writing it for each pair of parties and then summing the resulting inequalities, obtaining $\sum_{\alpha=1}^P\delta_\alpha\le P(N+1)/2$, see the discussion in section \ref{conc} of the main text. However, we can do better by using the following well-known lemma, the proof of which we include for completeness.
\begin{lem1}\label{lem1}
If operators $\{R_j\}_{j=1}^N$ span a space of dimension $\delta_R$, then with $Q_j\ne0~\forall{j}$, operators $\{R_j\otimes Q_j\}_{j=1}^N$ span a space of dimension no less than $\delta_R$.
\end{lem1}
\proof Suppose by contradiction that operators $\{R_j\otimes Q_j\}_{j=1}^N$ span a space of dimension $\hat\delta<\delta_R$. Then, ordering the set so the first $\delta_R$ of the $R_j$ are linearly independent, we have that there exist coefficients $c_{j}$ not all zero such that
\begin{align}\label{eqn93}
0=\sum_{j=1}^{\hat\delta+1}c_{j}R_j\otimes Q_j.
\end{align}
Linear independence of the set $\{R_j\}_{j=1}^{\hat\delta+1}$ appearing in this expression implies $c_j=0~\forall{j}$, a contradiction, completing the proof.\endproof

From this lemma, we have Theorem~\ref{thm106} of the main text.

{\bf Theorem~\ref{thm106}} \emph{Given a set of product operators for $P$ parties, $\{\MC_j=M_j^{(1)}\otimes\ldots\otimes M_j^{(P)}\}_{j=1}^N$, if there exists a set of nonzero coefficients, $\{c_j\}$, such that the linear combination $\SC=\sum_{j=1}^Nc_j\MC_j$ is a product operator, then for each pair of parties $\alpha,\beta$,
\begin{align}\label{eqn992}
\delta_\alpha+\delta_\beta\le N+1,
\end{align}
where $\delta_\alpha$ is the dimension of the space spanned by operators $\{M_j^{(\alpha)}\}_{j=1}^N$.}
\proof Consider a bipartite split $A|B$ of the $P$ parties, with party $\alpha$ on side $A$ and party $\beta$ on the other side, $B$. Then, by lemma~\ref{lem1}, $\delta_A\ge\delta_\alpha$ and $\delta_B\ge\delta_\beta$. The theorem follows directly, since $\delta_\alpha+\delta_\beta\le\delta_A+\delta_B\le N+1$, where the second inequality is just corollary \ref{cor2}.\endproof

Theorem~\ref{thm106} leads immediately to the following observation.
\begin{cor5}\label{cor5}
Every subspace spanned by a set of product operators is devoid of any other product operator unless there is a subset of the original set of product operators that satisfies \eqref{eqn992} of Theorem~\ref{thm106} for every pair of parties, $\alpha,\beta$.
\end{cor5}
\noindent The results for $r_s=1$ can be strengthened. Let $\widehat\delta$ be the dimension of the space spanned by product operators $\{\MC_j\}_{j=1}^N$. Then,
\begin{thm4}\label{thm4}If the set of product operators $\{\MC_j=\otimes_{\alpha=1}^PM_j^{(\alpha)}\}_{j=1}^N$ is linearly dependent so that $\widehat\delta<N$, and if $\SC=\sum_{j=1}^Nc_j\MC_j$ with $\SC$ a product operator, then for each pair of parties $\alpha,\beta$,
\begin{align}\label{eqn93}
\delta_\alpha+\delta_\beta\le\widehat\delta+1<N+1.
\end{align}
\end{thm4}
\noindent It is clear from this that we can simply replace $N$ by $\widehat\delta$ in both corollaries \ref{cor2} and \ref{cor5}.
\proof Choose indexing of the $\MC_j$ such that $\{\MC_1,\MC_2,\ldots,\MC_{\widehat\delta}\}$ are linearly independent. Then, for $j>\widehat\delta$, $\exists\{c_{ji}\}$ such that
\begin{align}\label{eqn112}
M_j^{(1)}\otimes M_j^{(2)}\otimes\ldots\otimes M_j^{(P)}=\sum_{i=1}^{\widehat\delta} c_{ji}M_i^{(1)}\otimes M_i^{(2)}\otimes\ldots\otimes M_i^{(P)},
\end{align}
which tells us that the linear combination on the right-hand side is a product operator. Hence, Theorem~\ref{thm106} immediately implies that for every pair $\alpha,\beta$, $\delta_\alpha^\prime+\delta_\beta^\prime\le\widehat\delta+1$, where $\delta_\alpha^\prime$ is the dimension of the span of the first $\widehat\delta$ of the $M_j^{(\alpha)}$, and similarly for $\delta_\beta^\prime$. However, from \eqref{eqn112}, multiplying by $M_j^{(2)\dag}\otimes\ldots\otimes M_j^{(P)\dag}$, and taking the partial trace over all parties except the first one, we see that $M_j^{(1)},~j>\widehat\delta,$ lies in the span of the first $\widehat\delta$ of the $M_i^{(1)}$, which implies that $\delta_1^\prime=\delta_1$. A similar argument shows that $\delta_\alpha^\prime=\delta_\alpha~\forall{\alpha}$, which completes the proof.\hspace{\stretch{1}}$\blacksquare$


\end{document}